\documentclass[twocolumn,aps,prl,superscriptaddress, amsmath,amssymb]{revtex4-2}

\usepackage{graphicx}
\usepackage{dcolumn}
\usepackage{bm}
\usepackage{epstopdf}
\usepackage{xcolor}
\usepackage{siunitx} 

\usepackage{amsmath} 

\usepackage{upgreek}  

\usepackage[normalem]{ulem} 

\newcommand{\Andrey}[1]{{\color{black} #1}}
\newcommand{\Bo}[1]{{\color{black} #1}}

\begin{document}

\title{Guiding self-assembly of active colloids by temporal modulation of activity}


\author{Bo Zhang}
\affiliation{Materials Science Division, Argonne National Laboratory, 9700 South Cass Avenue, Lemont, IL 60439, USA}

\author{Alexey Snezhko}
\affiliation{Materials Science Division, Argonne National Laboratory, 9700 South Cass Avenue, Lemont, IL 60439, USA}

\author{Andrey Sokolov}
\email{sokolov@anl.gov}
\affiliation{Materials Science Division, Argonne National Laboratory, 9700 South Cass Avenue, Lemont, IL 60439, USA}

\date{\today}


\begin{abstract}

Self-organization phenomena in ensembles of self-propelled particles open  pathways to the synthesis of new dynamic states not accessible by traditional equilibrium processes. The challenge  is to develop a set of principles that facilitate the control and manipulation of emergent active states. Here, we report that dielectric rolling colloids energized by a pulsating electric field self-organize into alternating square lattices with a lattice constant controlled by the parameters of the field. We combine experiments and simulations to 
examine spatiotemporal properties of the emergent collective patterns, and investigate the underlying dynamics of the self-organization.
We reveal the resistance of the dynamic lattices to compression/expansion stresses leading to a hysteretic behavior of the lattice constant.
The general mechanism of pattern synthesis and control in active ensembles via temporal modulation of activity can be applied to other active colloidal systems. 


\end{abstract}

\maketitle


The subject of vast interest in the field of active matter physics is the emergence of self-organized collective behavior out of chaotic motions of individual particles as a result of inter-particle interactions \cite{sanchez2012spontaneous, snezhko2011magnetic, palacci2013living, bastien2020model, zhang2020reconfigurable, karani2019tuning, sokolov2019emergence}. Self-organization phenomena in biological active systems, such as flocks of birds and schools of fish, rely on communications, visual monitoring,  sensing of individual positions and requires constant brain processing of collected information \cite{vicsek2012collective, feinerman2018physics, cavagna2014bird, lavergne2019group}. 
A collective motion in ensembles of simple organisms, such as  swimming bacteria, may arise only out of inter-particle steric and hydrodynamic interactions \cite{sokolov2007concentration, gyrya2010model, sokolov2012physical, marchetti2013hydrodynamics, lushi2014fluid, nishiguchi2018engineering, hamby2018swimming} 
that makes them a popular model system for the investigation of collective behaviors and self-organization phenomena.

Systems of synthetic self-propelled particles, energized by chemical reactions or electromagnetic fields, provide even better control over individual and collective dynamics of active units and, as a result,  complex out-of-equilibrium dynamics of such systems is the  subject of extensive research \cite{bricard2013emergence, sanchez2015chemically, ebbens2018catalytic, han2020reconfigurable, zhang2020reconfigurable, karani2019tuning, zhang2021active, bauerle2020formation, driscoll2017unstable, yan2016reconfiguring}. Currently, the majority of collective phenomena are observed in the systems with a constant (or nearly constant) energy injection rate. For instance, a system of electrostatically driven rolling colloids, Quincke rollers, exhibits a large variety of dynamic phases ranging from an isotropic gas to polar bands, vortices, swarms, and rotating clusters \cite{bricard2013emergence,bricard2015emergent,zhang2020reconfigurable,zhang2021quincke}. The temporal modulation of activity in that system was recently suggested to model behavior of living systems and control clustering of particles \cite{karani2019tuning,zhang2021persistence,zhang2021reversal}.

In this Letter, we report self-assembly of active Quincke rollers into dynamic square lattices when energized by a spatially uniform but modulated in time electric field. 
The formation of dynamic structures is triggered by a momentarily decoupling between dominant inter-particle interactions - hydrodynamic velocity alignment and near-field electrostatic repulsion as a result of a temporal cessation of the energy injection. Temporal cessation of the activity  resets particles' interactions and redirects velocities according to the locally formed particle arrangements. In contrast to vertically vibrated granular matter \cite{douady1989subharmonic,fauve1989collective,umbanhowar1996localized,melo1995hexagons},  only square lattices are observed, and these lattices preserve their structure after  termination of the activity.

\begin{figure*} [!htbp]
\centering
\includegraphics[width=0.85\linewidth]{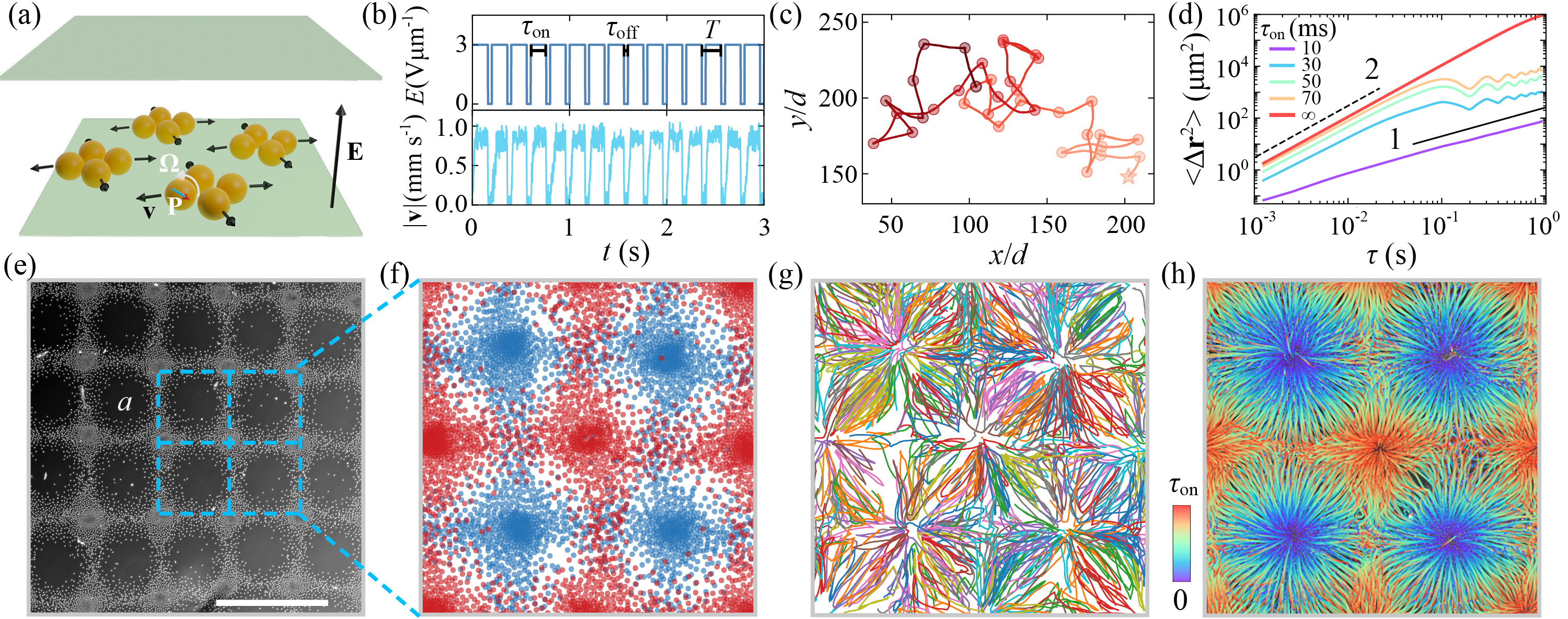}
\caption{
     Self-assembly of Quincke rollers into square lattices under a pulsating electric field. 
	(a) A sketch of the experimental setup. 
	(b) Temporal dependence of the electric field and the averaged particles' velocity. \Andrey{$T=\tau_\text{on}+\tau_\text{off}$ is the period of the signal}. 
	(c) A trajectory of an isolated roller  energized by the pulsating electric field. 
	(d) Mean square displacement curves of the rollers under periodic activity modulations.  
	The bottom (purple) and the top (red) curves correspond to the motion of the rollers in Gas and Vortex phases respectively. 
	 $E =$ 2.7 V $\SI{}{\micro\meter}^{-1}$ and $T=100$ ms for all curves except the case of a constant field.
	Slopes marked as 1 and 2  indicate diffusive $(\sim \tau)$ and ballistic $(\sim \tau^2)$ regimes respectively.
	(e) An experimental snapshot of a dynamic square lattice captured between electric field pulses.	The blue squares marks four unit cells with a lattices constant $a$. \Bo{The same area is also shown in (f-h)}. 
	The particle area fraction $\phi=$ 0.114; $E =$ 3.0 V $\SI{}{\micro\meter}^{-1}$; $T=125$ ms; $\tau_\text{on}=112.5$ ms. The scale bar is 0.5 mm. See also Video S1 \cite{SM}. 
	(f) \Andrey{Overlayed square lattices formed after two consecutive cycles represented by blue and red circles.}
	(g) \Andrey{Visualization of the particle trajectories over five periods of the field. For clarity only 10 \% of particle trajectories are shown.} 
	(h) \Bo{Color-coded visualization of the time evolution of particles positions during one cycle}. The color \Andrey{in each point of a particle trajectory corresponds to the time between 0 (purple) and $\tau_\text{on}$ (red) as shown by the colorbar}. 
	}
\label{Fig1}
\end{figure*}

In our experiments, spherical polystyrene particles of the diameter $d=$ 4.8 \SI{}{\micro\meter} are dispersed in 0.15 mol L$^{-1}$ AOT/hexadecane solution and  sandwiched between two parallel ITO-coated glass slides spaced 45 \SI{}{\micro\meter} apart, see Fig.~\ref{Fig1}(a). 
When a static (DC) electric field is applied, particles  polarize, and above the critical amplitude of the  field \citep{bricard2013emergence} they start to steadily rotate and roll on the bottom surface with a constant speed driven by electrohydrodynamic Quincke rotation phenomenon \cite{quincke1896ueber,tsebers1980internal}. \Andrey{The typical velocity of rollers  in our experiments is 0.8 mm/s.} 
At a particle area fraction above $\phi_\text{c} \approx 0.002$, rollers form a steady vortex or a traveling band in confined systems \cite{bricard2013emergence, bricard2015emergent, zhang2020oscillatory}. 
The temporal profile of the mean square displacement (MSD) for individual rollers (red curve in Fig.~\ref{Fig1}(d)) reveals long ballistic regime of motion.

The behavior of rollers becomes drastically different if activity of particles is modulated by the pulsating electric field shown in Fig.~\ref{Fig1}(b). For simplicity, we fix the magnitude of the electric field, but probe the rollers' dynamics by varying only the duration of pulses $\tau_\text{on}$ and the interval between the pulses $\tau_\text{off}$. This technique was recently suggested to model run-and-tumble behavior and L\'evy walks via incomplete depolarization of rollers at relatively short resting time $\tau_\text{off}$ \cite{karani2019tuning}. The formation of lattices observed in our work requires full depolarization of particles after each cycle of the electric field which occurs if $\tau_\text{off}$ is several times larger than the Maxwell-Wagner polarization relaxation time $\tau_\text{MW}$ ($\tau_\text{MW}=(\epsilon_\text{p}+2\epsilon_\text{f})/(\sigma_\text{p}+2\sigma_\text{f}) \approx 1$ ms), where $\epsilon_\text{p, f}$ and $\sigma_\text{p, f}$ are the permittivities and conductivities of particles (p) and fluids (f), respectively. Relaxation of \Andrey{large-scale} flows and hydrodynamic interactions \Andrey{occur within viscous timescale that in the experimental cell of the thickness $d=45$ \SI{}{\micro\meter} is of the order of 1 ms.} As a result, each roller in our experiments comes to a complete rest  for $\tau_\text{off}>10$ ms. \Andrey{The Brownian motion has a negligible effect on particles' motion due to their large size and does not alter their positions when activity is terminated.} At low concentrations ($\phi < \phi_c$), each particle randomizes the velocity direction in each period (Fig.~\ref{Fig1}(c) \Bo{and Fig. S9 \cite{SM}}).

At a high particle concentration, the initial direction of motion upon field application is affected by an arrangement of neighbors. Rollers tend to move against the local gradient of particle density due to the electrostatic repulsion between polarized particles \cite{zhang2021reversal}. The spontaneous formation of a cluster is followed by its quick explosive decay and concentration of particles in previously depleted areas during the next cycle. 
While initially the positions of nuclei clusters are random, they slowly evolves into a well-defined stable structure, and in several hundred activity cycles particles' positions start to alternate between two sets of nearly perfect square lattices ($\bf{A}$ and $\bf{B}$) with the same lattice constant $a$, see Fig.~\ref{Fig1}(e,f), Fig. S1, and Video S1 \cite{SM}. 
The structure of lattices $\bf{A}$ and $\bf{B}$ are geometrically identical but translated by the half of a diagonal of the unit cell ($a/2$, $a/2$). 
\Bo{A closer inspection of particles' motion shows that the average distance particles travel each cycle is noticeably smaller than the shift between the lattices ($\sqrt{2}a/2$), \Andrey{see Fig.~\ref{Fig1}(g,h).}}  
The particles preserve their positions in one of two lattices as long as the field is off, but quickly form the alternative lattice upon the next pulse of the electric field. Therefore, the observed self-organization of particles is intrinsically different from activity-induced cluster formation \Andrey{in which active particles slow down in crowded areas, further increasing the local density and the size of the cluster.} \cite{kagan2011chemically, buttinoni2013dynamical, palacci2013living}.
The reciprocating motion of rollers between the two lattices (period is $2T$) effectively traps individual particles within unit cells for several cycles, as reflected by the oscillation of MSD curves, and particle trajectories (Fig.~\ref{Fig1}d,g). The motion of particles eventually becomes diffusive as particles migrate from cell to cell over the system. The diffusion constant $D$ increases with particle run time $\tau_\text{on}$, see Fig. S3(a) in \cite{SM}. 

\begin{figure} 
\centering
\includegraphics[width=1\linewidth]{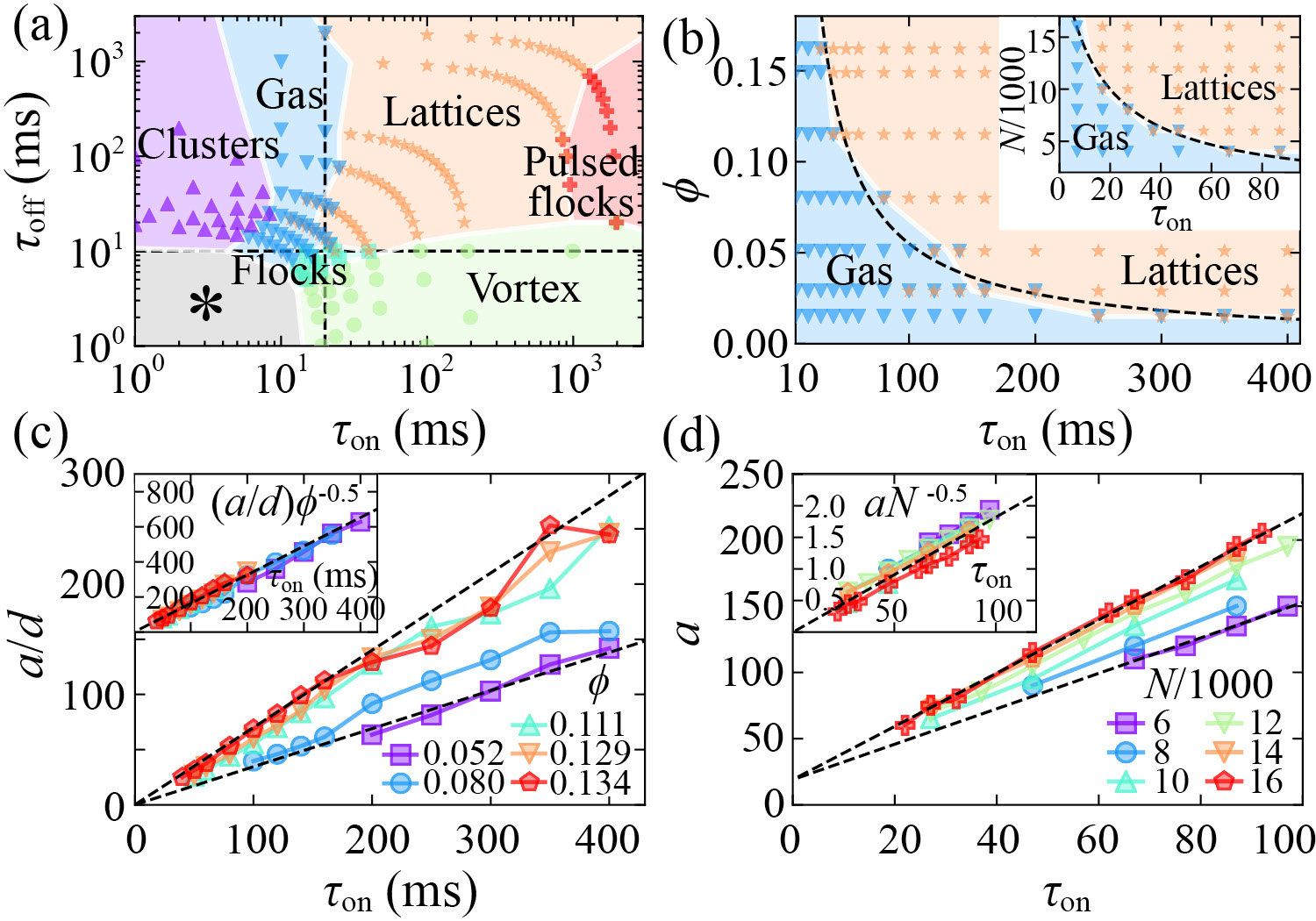}
\caption{
	(a) Dynamic phases formed by rollers for different $\tau_\text{on}$ and $\tau_\text{off}$. $\phi=$ 0.120; $E =$ 2.7 V $\SI{}{\micro\meter}^{-1}$. The system is confined by a cylindrical well with a diameter $D=$ 1 cm. The phases marked as "$*$"  have been previously reported in Ref.~\cite{karani2019tuning}.  See Video S3 in \cite{SM} for more details. 
	(b) Lattices and Gas formed at different $\tau_\text{on}$ and $\phi$ ($\tau_\text{off}\ge$ 40 ms). The dashed line is a fit of the boundary between phases by a function $\phi = A / \tau_\text{on}$, see text for details. \Bo{Inset}: dynamic phase diagram obtained from simulations. $N$ is the number of particles in the system. 
	(c-d) Dependence of the lattice constant on $\tau_\text{on}$ for different area fractions in experiments (c) and simulations (d).
	$D=$ 1.5 mm. 
	The dashed lines are linear fits for the lowest and highest area fractions. \Bo{Insets}  in (c) and (d)  experimental and simulation data with the lattice constant re-scaled by a $\sqrt{\phi}$. 
	}
\label{Fig2}
\end{figure}

Our minimalistic phenomenological model (see \cite{SM} for details) that involves only isotropic inter-particle repulsion and velocity alignment mechanisms accurately reproduces the formation of patterns observed in experiments (see Fig. S2, S3 in \cite{SM}) as well as main dynamic properties of the lattices discussed below.  
The results of our simulations highlight the role of repulsive forces immediately upon the system activation. In the model, while the system is active all particles move with nearly the same constant speed (small fluctuations due to noise or interactions do not alter the results). The initial direction of motion for each particle at the beginning of each cycle is defined by the net repulsive force from the neighbors. If the initial direction were  chosen randomly, particles would not form any stable structure. These findings provide a hint on the mechanism of  lattice formation. Upon the system activation, particles start to experience repulsion from the neighbors (due to a field-induced polarization) and roll away from dense clusters. A temporal cessation of the activity and subsequent re-energizing of the system lead to rapid  reorientation of particles' velocities against the local density gradients, and results in a reciprocating motion of particles between the two sets of lattices.  



In order to gain additional insights into the observed self-organization, we explore the response of the system to changes in the pulse duration, $\tau_\text{on}$, and distance between the pulses, $\tau_\text{off}$, see  Fig.~\ref{Fig2}a \Bo{and Video S3 in \cite{SM}}.
The lower limit of $\tau_\text{on} \approx$ 20 ms for the lattices phase comes from a minimum time and distance rollers must travel to interact with each other as well as to overcome the intrinsic positional noise in the system. This limit naturally depends on the  particles density and may be significantly longer for low densities as discussed below. 
For $\tau_\text{on}>$ 1 s the particles' velocities at the end of each cycle become uncorrelated with initial orientations and, therefore, rollers are not able to alternate between fixed stable patterns. 
The vortex phase in Fig.~\ref{Fig2}(a) corresponds to the formation of a continuous single vortex at $\tau_\text{off}<$ 10 ms. \Andrey{At this phase, the termination of energy injection is so brief that particles do not fully stop.} In the pulsed flocks phase the formation of the global vortex is interrupted every period of the signal. At small $\tau_\text{on}$ rollers' dynamics resembles a gas motion with a tendency to form clusters as  $\tau_\text{on}$ further decreases, see Fig.~\ref{Fig2}(a). \Andrey{The Lattices can be distinguished from the Gas by periodic oscillations of the density profile, see Fig. S4 and S6 in \cite{SM}. The crossover between those two dynamic states is continuous and smooth (see Fig. S4).} Complex behavior of the system  at small $\tau_\text{off}$ and $\tau_\text{on}$ (marked gray in Fig.~\ref{Fig2}(a)), corresponds to  rollers with incomplete depolarization studied in  \cite{karani2019tuning}.

Similar to other active systems, the formation of globally correlated states (like lattices) can only be observed at particle densities  above a certain threshold. We investigate in experiment and simulations the behavior of the critical particle number density, necessary to facilitate the formation of lattices, as a function of the run time, $\tau_\text{on}$. \Andrey{This transition is accessible only at $\tau_\text{off}$ and $\tau_\text{on}$ time larger than the polarization relaxation time $\tau_\text{MW}$.} At low densities particles need longer time to reach and interact with the neighbors, and as a result, a longer run time is required to support the dynamic lattice. Correspondingly, as the density of the rollers increases, the minimum run time to form the lattice decreases, see Fig.~\ref{Fig2}(b). \Andrey{The boundary between the Gas and Lattices follows a simple relation connecting the particle number density, $\phi$, and the run time $\tau_\text{on}$: $\phi= A/\tau_\text{on}$. The above scaling is valid for the whole range of scanned parameters (10 ms $<$ $\tau_\text{on}$ $<$ 400 ms and 0.015 $<$ $\phi$ $<$ 0.17), and $A$ here is a constant that depends on the properties of the system (particle size, activity, and liquid media). $A=$ 5.5 ms for our experimental systems. }

\Andrey{The lattice constant $a$ growths with $\tau_\text{on}$, see Fig.~\ref{Fig2}c. Nevertheless, all the lines obtained at different densities collapse into a single line when re-scaled by a $ \sqrt{\phi}$, see Fig.~\ref{Fig2}c inset.} 
\Andrey{Such scaling possibly comes from the similarity of the lattice density profiles at different $\phi$ that scale with the size of the unit cell, see Fig. S12 in \cite{SM}.} Correspondingly, the evolution of the lattice constant $a$ with the run time $\tau_\text{on}$ can be written as $a= k d \tau_\text{on} \sqrt{\phi}$, where $k$ is the system dependent constant, \Andrey{the slope of the re-scaled line} ($k$ = 1.63 ms$^{-1}$ for our system).  The simulation results capture the similar trend, see Fig.~\ref{Fig2}(d). \Andrey{The above scalings can be used to estimate a number of particles per unit cell of the lattice at the transition point between the gas and lattice phases.} As the particles' distribution over the system is homogeneous and, therefore, the number of particles per unit cell of the square lattice can be estimated as $n=4 a^2 \phi/ (\pi d^2)$. Substituting $a$ and $\phi$ by its dependence on $\tau_\text{on}$ \Andrey{at the transition boundary between the gas and lattice phases}, one obtains that the number of rollers per unit cell at the transition is $n_\text{c}=4 k^2 A^2/ \pi $ (see more details in Note S2) \cite{SM}. 
\Andrey{Since $n_c$ is $\phi$ and $\tau_{on}$ independent, the number of particles per unit cell is the same at any transition point between the gas and lattice phases (the dashed line in Fig. 2b) regardless of the size of the unit cell.}

\begin{figure} 
\centering
\includegraphics[width=1\linewidth]{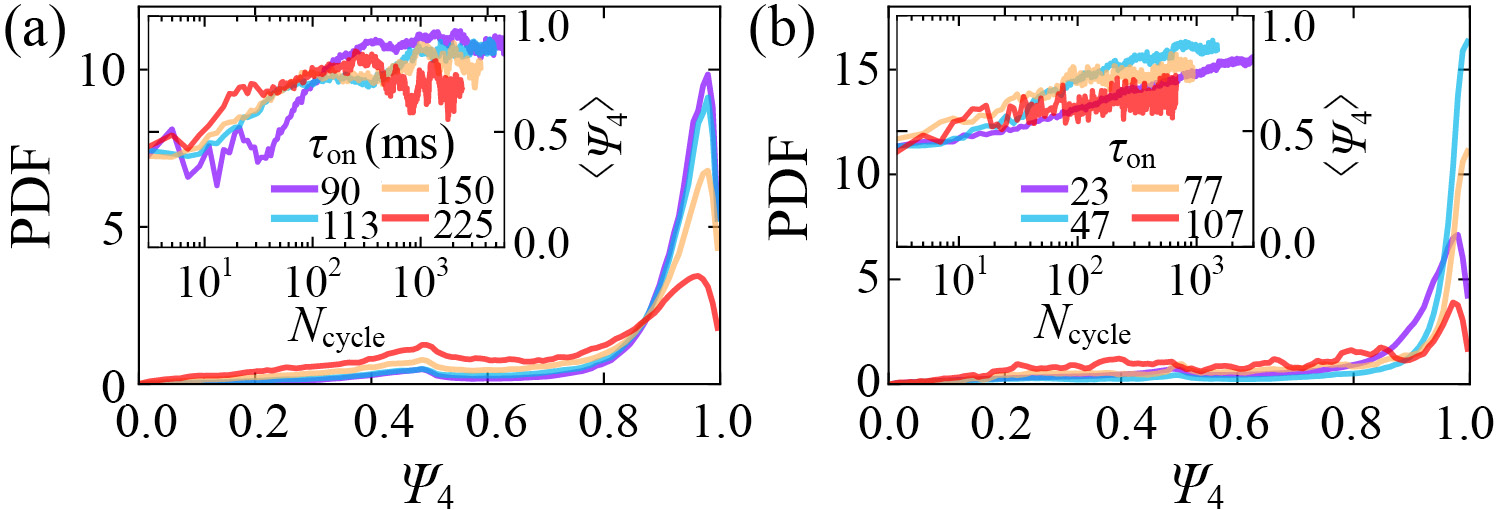}
\caption{
	Probability distribution functions (PDFs) of the local order parameters $\Psi_4$ in a stable lattice formed at different $\tau_\text{on}$ in experiments (a) and simulations (b). 
	\Bo{Insets}: Temporal evolution of  $\langle\Psi_4\rangle$. 
	$E =$ 3.0 V $\SI{}{\micro\meter}^{-1}$; $\tau_\text{on}=$ 112.5 ms; $\tau_\text{off}=$ 12.5 ms; $\phi$ = 0.114; $D=$ 1 cm.
	Also see Video S4-S5 in \cite{SM}.  
}
\label{Fig3}
\end{figure}

The dynamic square lattices do not have an ideal order at macroscopic distances and often develop a number of defects, see Fig. S2 \cite{SM}. To characterize the order of self-organized lattices we calculate the local degree of 4-fold symmetry for each vertex $j$ in a lattice: $\Psi_{4}=1/4 \vert \sum_{k=1}^{4}{\textrm{exp}(4i\theta_k)} \vert$, where $\theta_{k}$ are polar angles of four closest neighbors with the origin at $j$ vertex.
The probability distribution functions (PDFs) of $\Psi_4$ for stable lattices at different $\tau_\text{on}$ show peaks at $\Psi_4=1$, indicating a presence of a well defined  4-fold symmetry in the order, see Fig.~\ref{Fig3}. 
The peaks get suppressed with the increase of the run time $\tau_\text{on}$, reflecting the emergence of imperfections due to fluctuations of particles' velocities and the development of a global collective motion (vortex).
The formation of the \Bo{local} order may be further quantified by a time evolution of the order parameter $\langle \Psi_4 \rangle$ defined as $\Psi_4$ averaged over the whole system, see Fig.~\ref{Fig3}(a-b) \Bo{insets}.
The temporal evolution of $\langle \Psi_4 \rangle$ suggests that dynamic lattices emerge within a hundred cycles, however, they never become geometrically perfect as $\langle \Psi_4 \rangle$ reaches the plateau ($\approx0.8$). 
The system does not fully anneal all the defects even after many thousands of activity cycles.


\begin{figure} 
\centering
\includegraphics[width=1\linewidth]{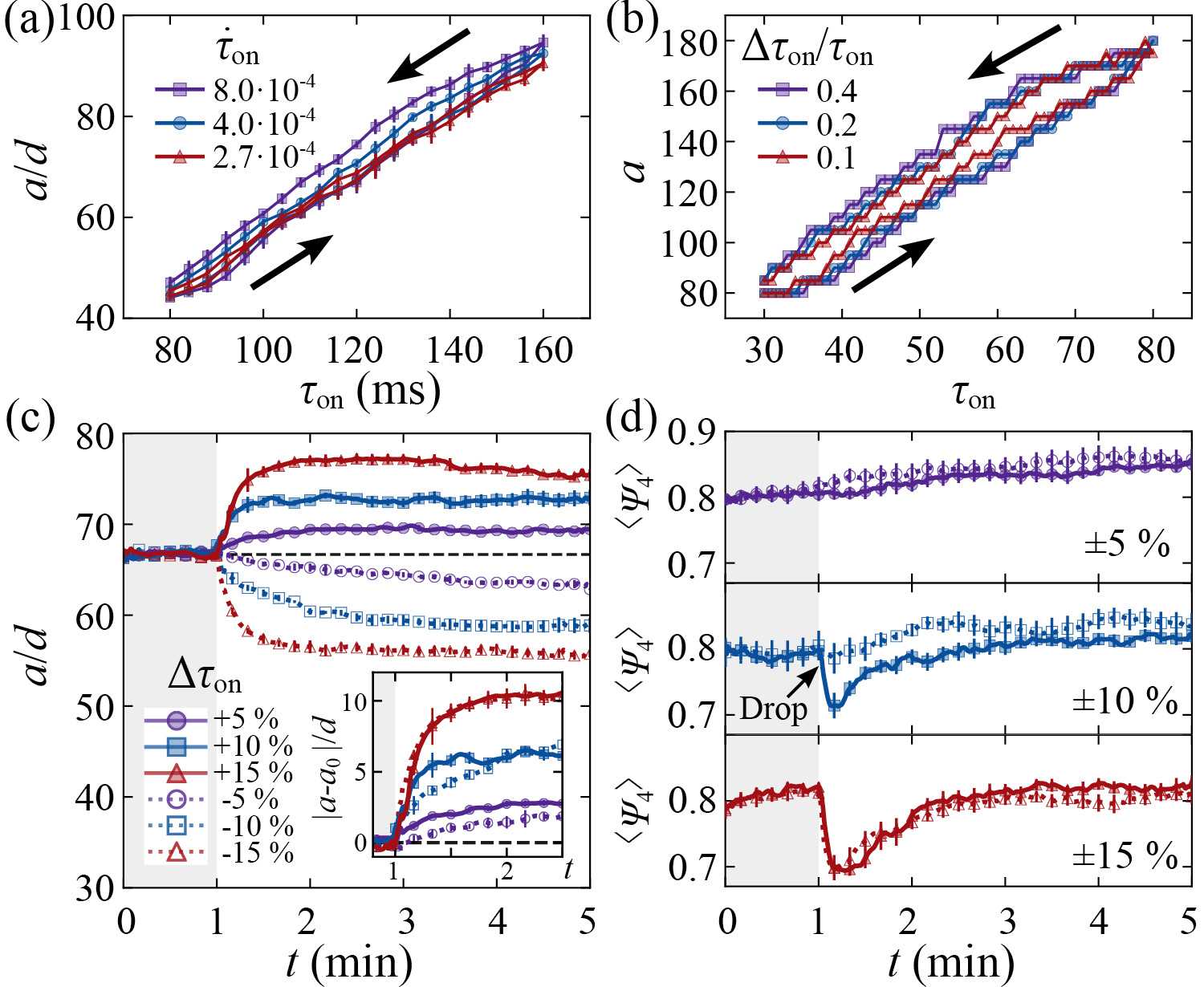}
\caption{
    Hysteresis loops for different ramping rates of $\tau_\text{on}$ in experiments (a) and simulations (b).  
    Arrows indicate the direction of the hysteresis loop. The starting point is $\tau_\text{on}$ = 80 ms (experiments) and 30 (simulations). See also video S6 and S7 \cite{SM}.
	(c-d) The evolution of the lattice constant (c) and the local order parameter $\langle\Psi_4\rangle$ of the square lattices (d) in response to the increase (solid symbols) or decrease (open symbols) of the $\tau_\text{on}$ 
	relative to the initial value of $\tau_\text{on}=$ 120 ms.
	The data are averaged over 5 experimental realizations for each curve.
	\Bo{Inset} in (c) shows the absolute change of $a$ compared to the initial value $a_0$.
}
\label{Fig4}
\end{figure}

In Quincke rollers system, the mechanism of self-organization \Andrey{with two alternating complementary lattices regulated by a density gradient} limits the types of stable lattices to square lattices for any combination of excitation  field parameters, while granular vibrated systems  often demonstrate transitions between stripes, square, and hexagonal lattices with a change of the driving parameters\citep{bizon1998patterns, aranson2006patterns,douady1989subharmonic,fauve1989collective,umbanhowar1996localized, melo1995hexagons}. 
The fixed  geometry of the lattices at a wide range of the driving field parameters enables the study of the collective memory effects \citep{couzin2002collective} in the system. Ramping the control parameter $\tau_\text{on}$ up and down at a fixed rate reveals a dynamic hysteresis on $a$ vs $\tau_\text{on}$, see Fig.~\ref{Fig4}(a-b). The size of the hysteresis loop increases with the ramping rate. 
Interestingly, the lattices  demonstrate asymmetry in the resistances against the expansion and compression stresses at a certain  range of stresses, see Fig.~\ref{Fig4}(c-d). The lattices smoothly transition to a smaller size unit cells if  $\tau_\text{on}$ is decreased by 10~\%, but break down and reassemble into a larger lattices when $\tau_\text{on}$ is increased by the same amount. This is manifested by a temporal drop of the order parameter, see Fig~\ref{Fig4}(d). If the relative change of $\tau_\text{on}$ is larger than 15~\% then lattices prefer to completely disintegrate for both compression and expansion stresses. In addition, the response of the lattices to the compression occurs faster than to expansion if  $\tau_\text{on}\leq$ 10~\%, see Fig~\ref{Fig4}(c).


In conclusion, the temporal modulation of activity in the system of Quincke rollers provides a robust technique for accessing and controlling dynamic self-organized states that are not available upon continuous energy injection.   A combination of experiments and numerical simulations has been used to investigate the physical mechanism that guides the formation of the reported patterns. A dominant role of electrostatic repulsion over hydrodynamic velocity alignment interactions immediately upon system reactivation results in a reciprocating motion of rollers between two stable square lattices. The lattices are re-configurable with the control of the characteristic lattice constant by the particles' run time.  The minimal number of particles per unit cell of the lattice required to facilitate the lattice formation is independent on the lattice constant and the average particle number density.
Our results provide new insights into the collective behavior and control of active colloidal ensembles by means of a temporal modulation of activity. The reported mechanism should, in principle, be applicable to other active systems where the collective behavior is governed by the interplay of isotropic repulsion and hydrodynamic velocity alignment interactions.

\begin{acknowledgments}
The research was supported by the U.S. Department of Energy, Office of Science, Basic Energy Sciences, Materials Sciences and Engineering Division. Use of the Center for Nanoscale Materials, an Office of Science user facility, was supported by the U.S. Department of Energy, Office of Science, Office of Basic Energy Sciences, under Contract No. DE-AC02-06CH11357.
\end{acknowledgments}

\bibliographystyle{apsrev4-2}  


%

\end{document}



\date{}
\maketitle

\clearpage


\begin{figure}
\centering
\includegraphics[width=1.0\textwidth]{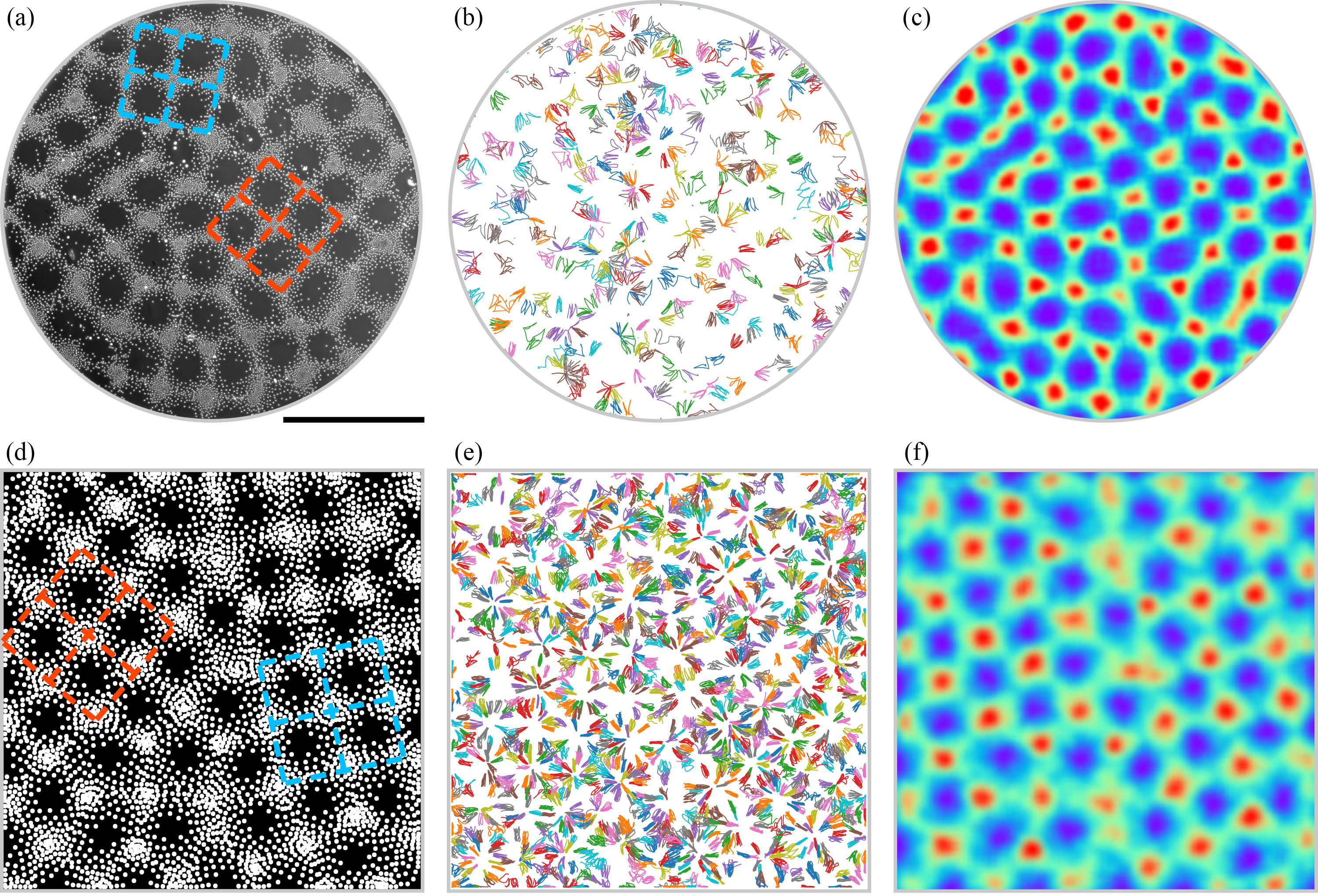}
\caption{
	Experimental (a-c) and simulation (d-e) results of lattices with multiple domains.
	(a, d) Typical images of the lattices. The blue and red squares mark unit cells in two different domains. See pattern switching in Movie S1, S2. The scale bar is 0.5 mm. 
	(b, e) Particle trajectories of the rollers over 10$T$. Only 5 \% and 30 \% of particle trajectories are shown, respectively.
	(c, f) Particle density maps of images shown in (a, d). The system is divided into smaller squares with sizes of 40 pixels (10.4$d$) for experimental images or 20 pixels for simulation images and with 80\% overlaps.  The densities are average values within squares.
}
\label{FigS1}
\end{figure}

\begin{figure}
\centering
\includegraphics[width=1.0\textwidth]{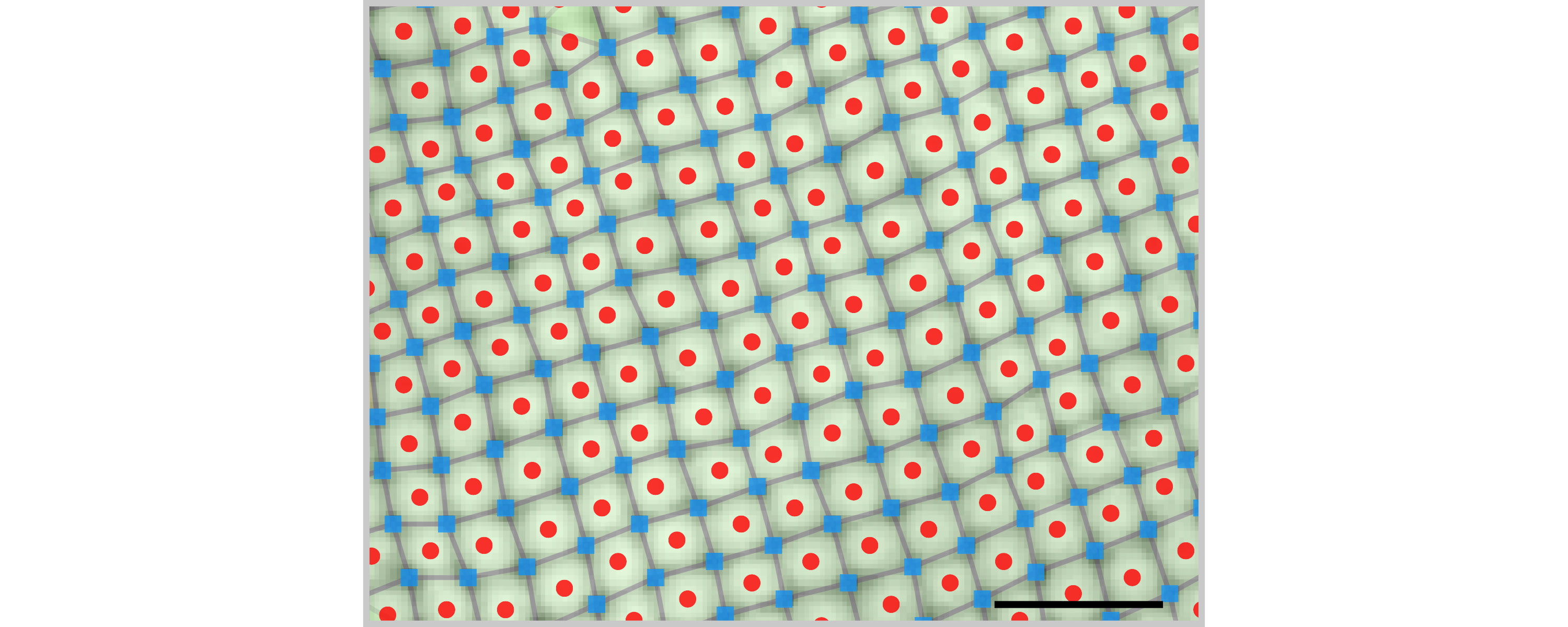}
\caption{
	Two sets of square lattices (vertices marked in blue squares and red circles respectively) in experiments with a time interval of $T$ in a large area (over 100 squares). $T$, $\tau_\text{on}$ and $\tau_\text{off}$ are 100 ms, 80 ms and 20 ms, respectively. The particle area fraction $\phi=$ 0.120. The scale bar is 0.5 mm. 
	The vertex positions are defined as local maxima of a particle density distribution. The vertex positions are further used for calculations of pair correlation functions and local order parameters. 
}
\label{FigS2}
\end{figure}

\begin{figure}
\centering
\includegraphics[width=1.0\textwidth]{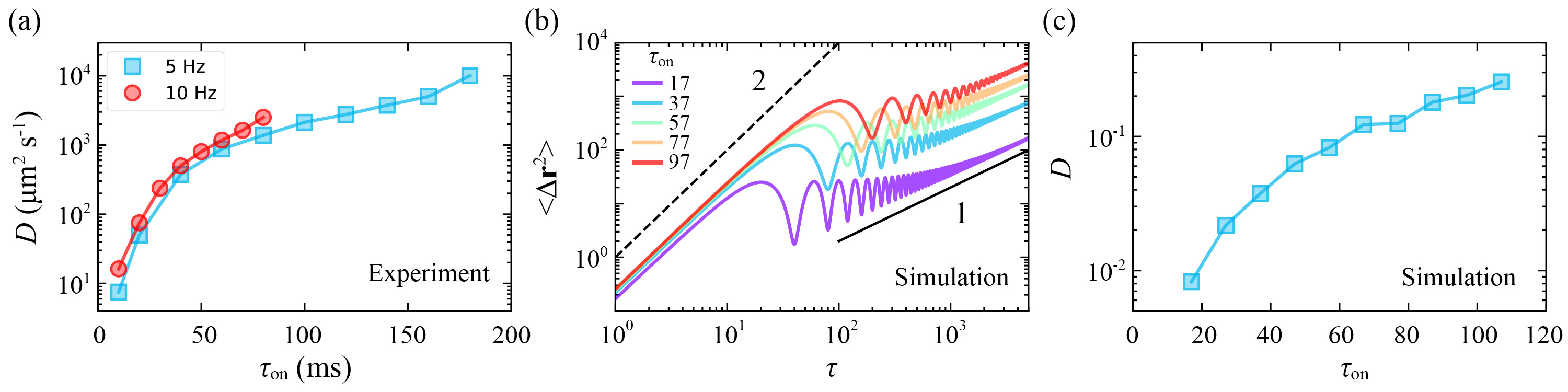}
\caption{
	Characterizations of particle dynamics of the lattices.
	(a) Diffusion coefficient constant $D$ of rollers increases with the field-on time $\tau_\text{on}$ in experiments. $D$ is measured based on the long time side of MSD curves where particle dynamics become diffusive as shown in Fig. 1d in the main text.  $\langle \Delta \textbf{r}^2 \rangle = 4Dt$. 
	The periods $T$ are 0.1 s (red circles with lines) and 0.2 s (blue squares with lines). 
	(b) Mean square displacements (MSDs) of particles in simulations. The oscillation of the curves in the long time indicates the back-and-forth motion of rollers as shown in Fig. 1i. The black solid line has a slope of 1 and the black dash line has a slope of 2.
	(c) $D$ of rollers increases with the field-on time $\tau_\text{on}$ in simulations. 
}
\label{FigS3}
\end{figure}

\begin{figure}
\centering
\includegraphics[width=1.0\textwidth]{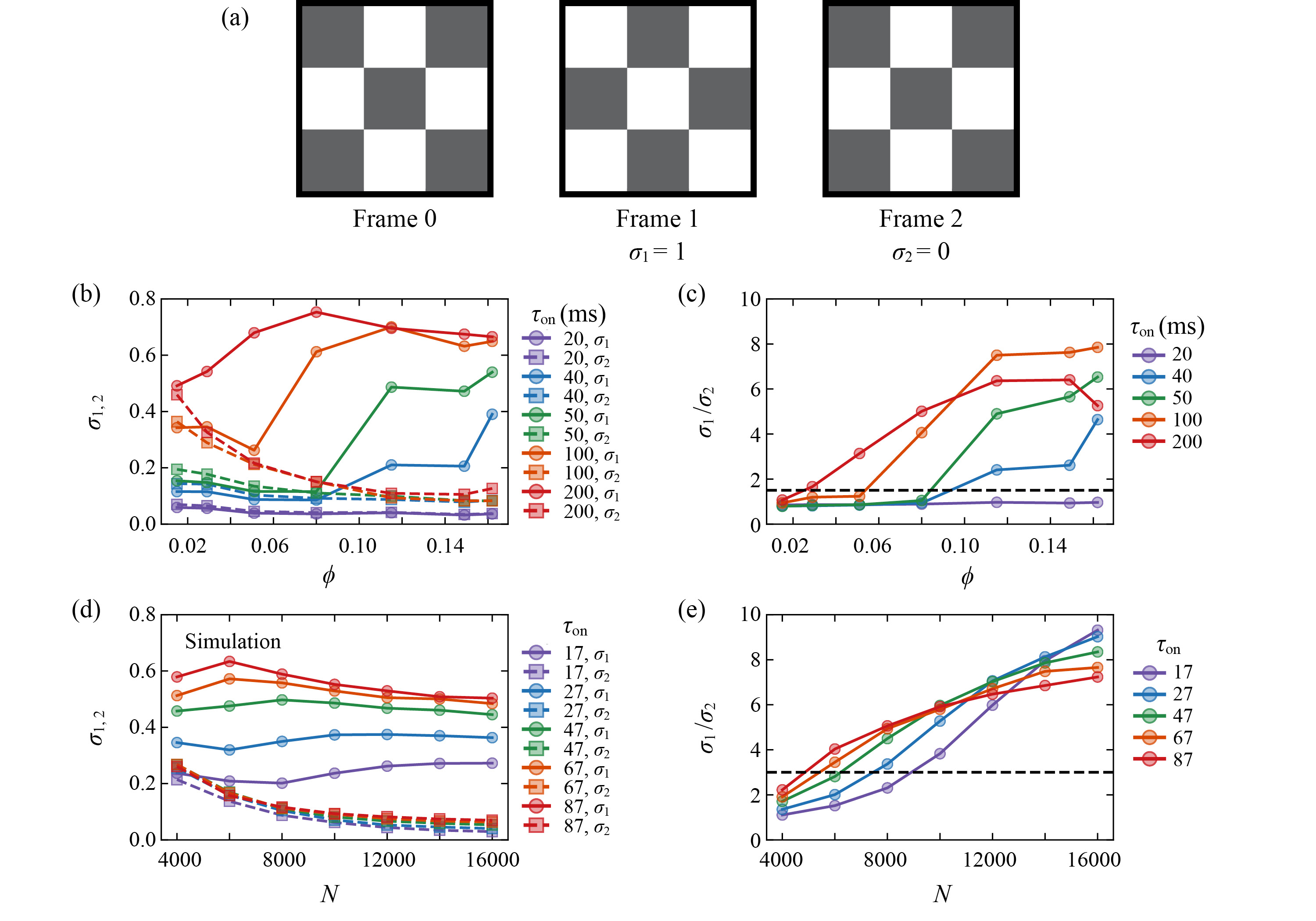}
\caption{
	Phases distinguished by density variations between frames. 
	(a) Schematics of an ideal density map for 3 successive stable frames which simplifies the density map of real lattices. The binary color represents the local area fraction $\phi_i$ in squares. Since there are two sets of lattices which switch back-and-forth, Frame 0 and 1 are complementary while Frame 0 and 2 are identical. 
	The density variation $\sigma$ between frames is defined as $\sigma_{1,2} = \sum_{i}|\Delta \phi_{i}| / (2N\langle \phi \rangle)$, where $\Delta \phi_{i}$ is the area fraction difference at the same square $i$ between two adjacent frames ($\sigma_{1}$) or two every other frames ($\sigma_{2}$). For an ideal density map (binary area fractions), $\sigma_{1}=1$ and $\sigma_{2}=0$. For real density map, the values of $\sigma_{1,2}$ are lower due to a broader distribution of area fractions. 
	(b-e) The density variation $\sigma_{1,2}$ (b, d) and the ratio $\sigma_{1}/\sigma_{2}$ (c, e) of the density maps with different field-on time and area fractions in experiments (b, c) and simulations (d, e). For lattices, $\sigma_{1}$ is large and $\sigma_{2}$ is small. The dash lines in (c, e) distinguish the Gas (below lines) and Lattices (above lines) phases. 
\Bo{In experiments, for convenience, we directly treat the brightness (inversed) of microscopic images (with some downsize and smooth) as density. While in simulations, the density map is calculated from a coarse-grained approach based on actual individual particle positions.}
}
\label{FigS4}
\end{figure}

\begin{figure}
\centering
\includegraphics[width=1.0\textwidth]{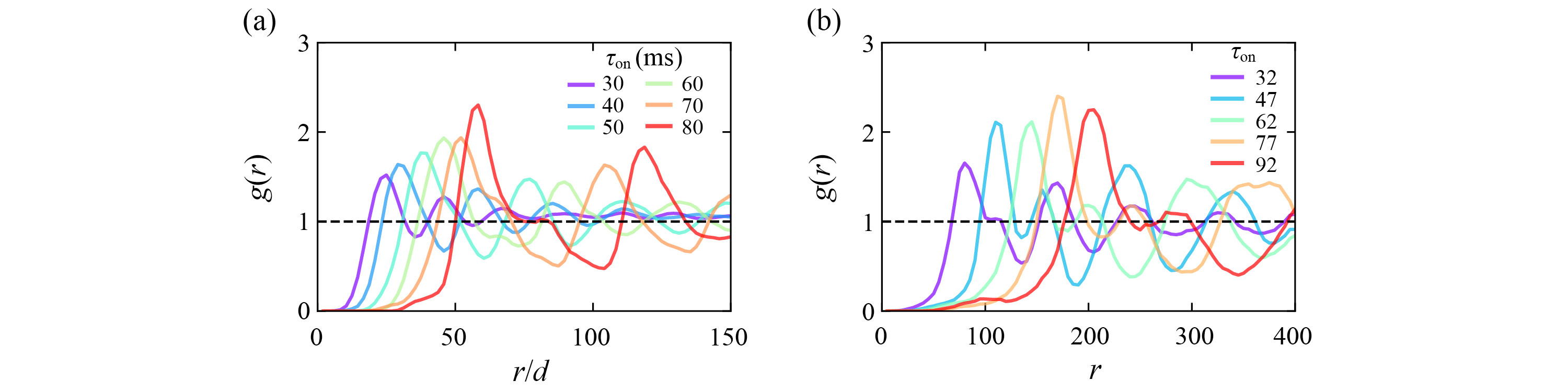}
\caption{
	Pair correlation functions $g(r)$ of lattices formed by rollers with different field-on time in experiments (a) and simulations (b). The period $T=0.1$ s for all experimental curves.
	The position of the first peak of $g(r)$ represent the lattice constant $a$. 
	}
\label{FigS5}
\end{figure}

\begin{figure}
\centering
\includegraphics[width=1.0\textwidth]{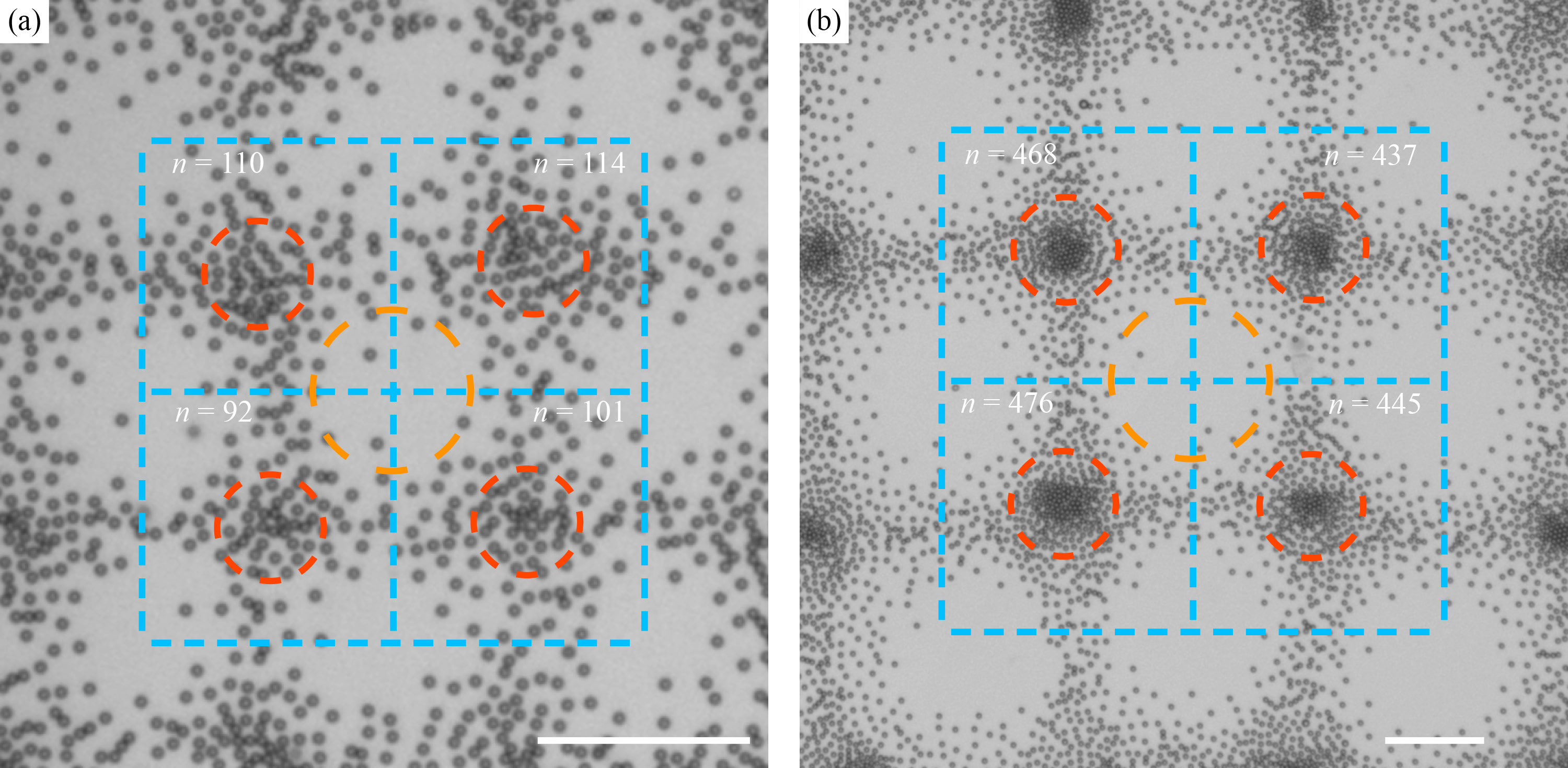}
\caption{
	The core-shell structures of unit cells for samples near (a) or away (b) from the Gas-Lattices phase boundary shown in Fig. 2b. The particle number $n$ in each unit cell is provided. The blue squares mark the unit cells; the red circles mark the dense cores; the orange circles mark the empty space. The loose particles between the red and orange circles behave like shells. The core-shell structures build a reliable particle density gradient which ensures a stable back-and-forth switch between two sets of patterns with a shift of (0.5$a$, 0.5$a$). The field-on time $\tau_\text{on}$ and the particle area fraction $\phi$ are 50 ms and 0.115 in (a) and 100 ms and 0.115 in (b), respectively. The number of particles of unit cells in (a) confirms the Eq.~\ref{n2}  ($n>$ 100). With the same area fraction but $\tau_\text{on}$ doubled in (b), the lattice constant doubles and the particle number in each unit cell increases to 4 times compared to the sample in (a).
The scale bars are 0.1 mm. 
}
\label{FigS6}
\end{figure}

\begin{figure}
\centering
\includegraphics[width=1.0\textwidth]{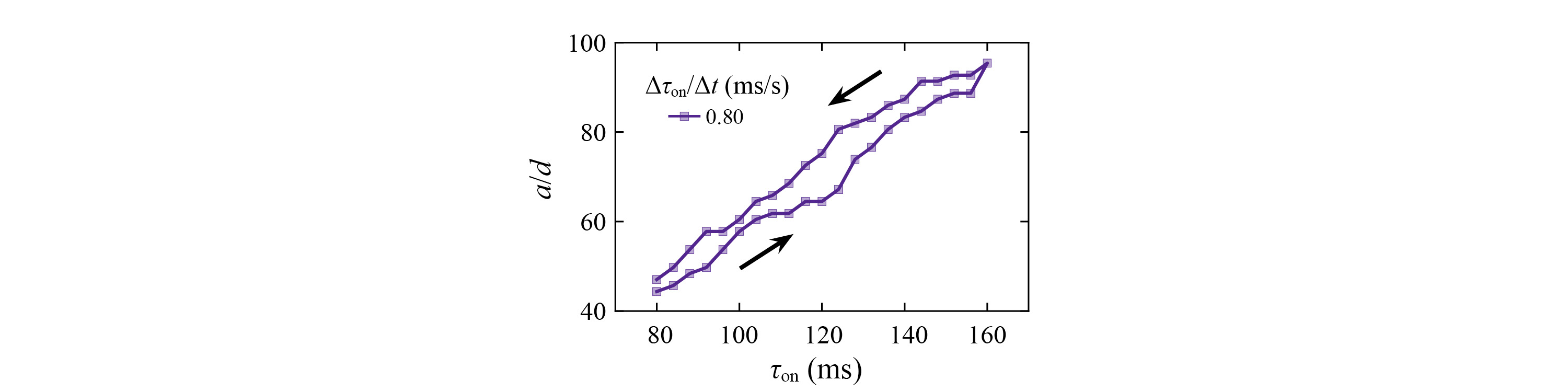}
\caption{
	Hysteresis of the lattice constant in a single measurement. 
}
\label{FigS8}
\end{figure}

\begin{figure}
\centering
\includegraphics[width=1.0\textwidth]{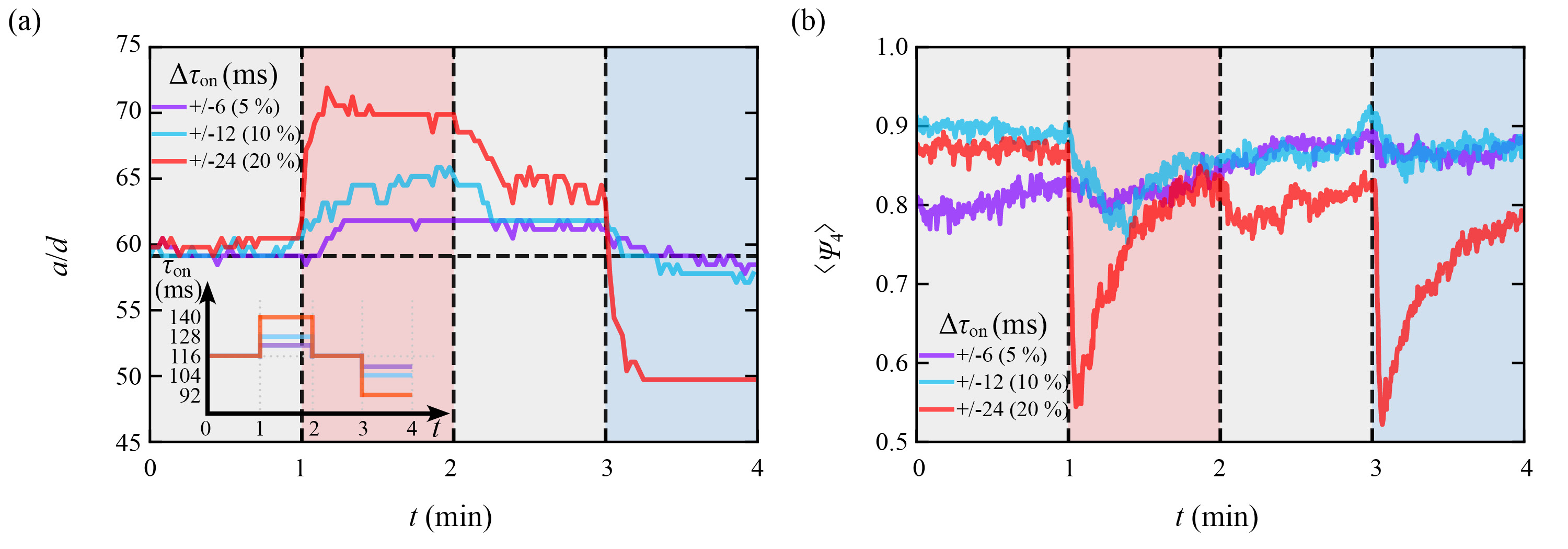}
\caption{
	The evolution of square lattices due to abrupt changes of $\tau_\text{on}$. 
	(a) The variations of the lattice constant $a$ of square lattices. The larger variation of $\tau_\text{on}$ yields larger variation of the lattice constant. The differences of $a$ between the grey bands confirm the hysteresis shown in Fig. 4 in the main text. \Bo{Inset}: The variation of $\tau_\text{on}$.
	(b) The local order parameter $\langle|\Psi_4|\rangle$ of square lattices. Square lattices can stand against small variations of $\tau_\text{on}$. Square lattices collapse (abrupt drop of $\langle|\Psi_4|\rangle$) and reassemble under large variations of $\tau_\text{on}$. Large changes at $t=$ 1 min and small changes at $t=$ 2 min and 3 min suggest that the lattices have a better tolerance of the compression compared to the expansion. Lattices collapse when $\tau_\text{on}$ changes further at $t=$ 3 min.
}
\label{FigS9}
\end{figure}

\begin{figure}
\centering
\includegraphics[width=1.0\textwidth]{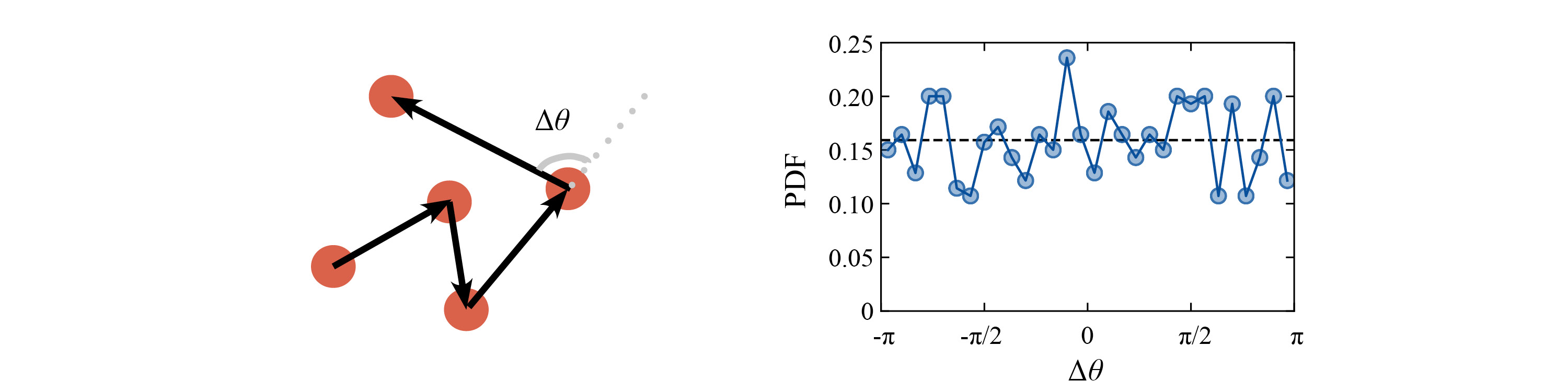}
\caption{
	Probability Distribution function (PDF) of the angle change $\Delta \theta$ of velocity direction of particles after each period. The PDF is calculated based on more than 600 events. The small fluctuations are statistical noise due to the relative small amount of counting events. The dash line indicates the average value of 1/(2$\pi$).
}
\label{FigS10}
\end{figure}

\begin{figure}
\centering
\includegraphics[width=1.0\textwidth]{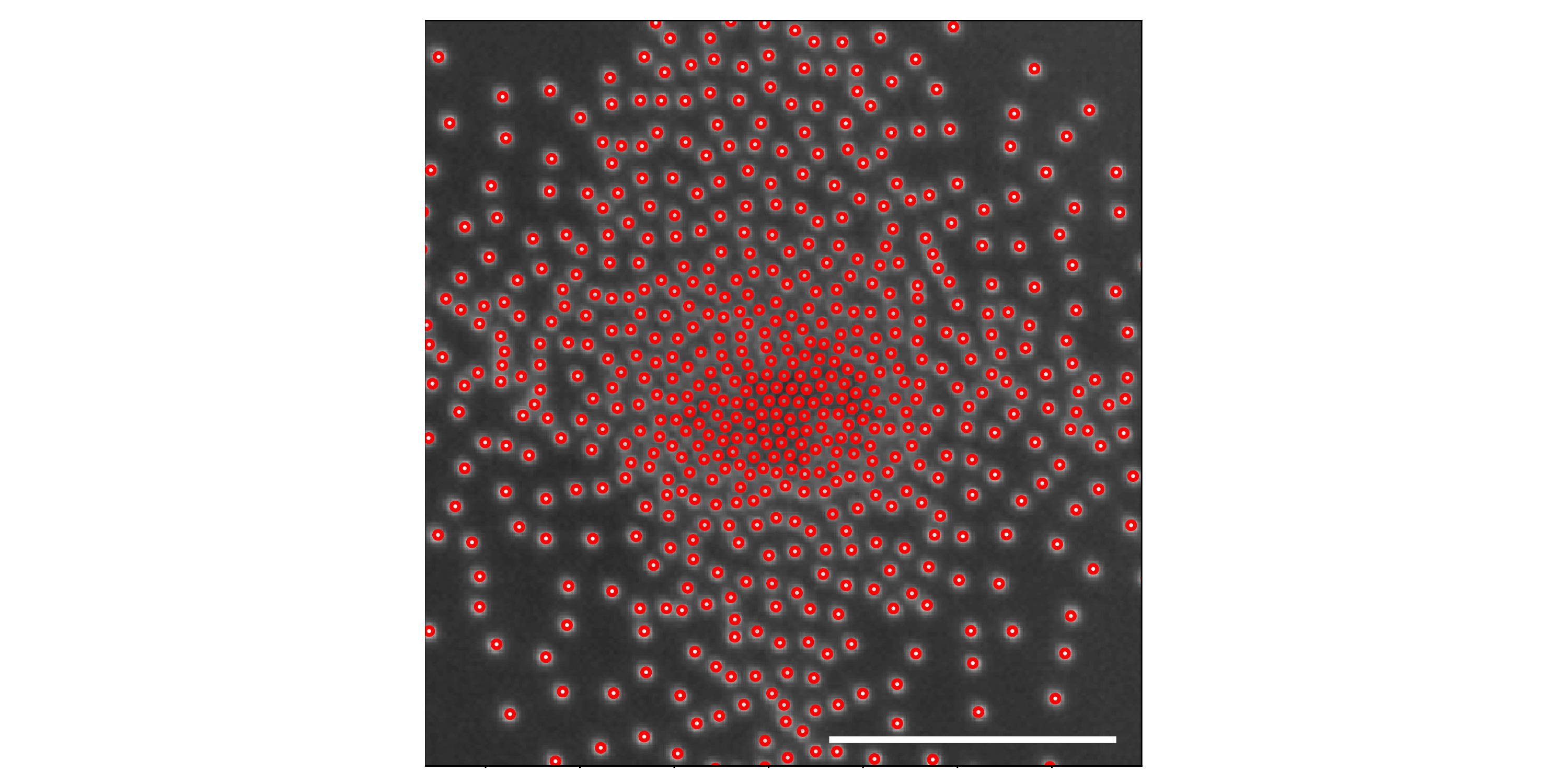}
\caption{
	Accurate feature locating of a monolayer of particles (marked in red circles) in one dense core of lattices. The scale bar is 0.1 mm. Particle tracking velocimetry and further data analysis are carried out with custom codes in Python and Trackpy \cite{allan2016trackpy}. 
}
\label{FigS11}
\end{figure}

\begin{figure}
\centering
\includegraphics[width=1.0\textwidth]{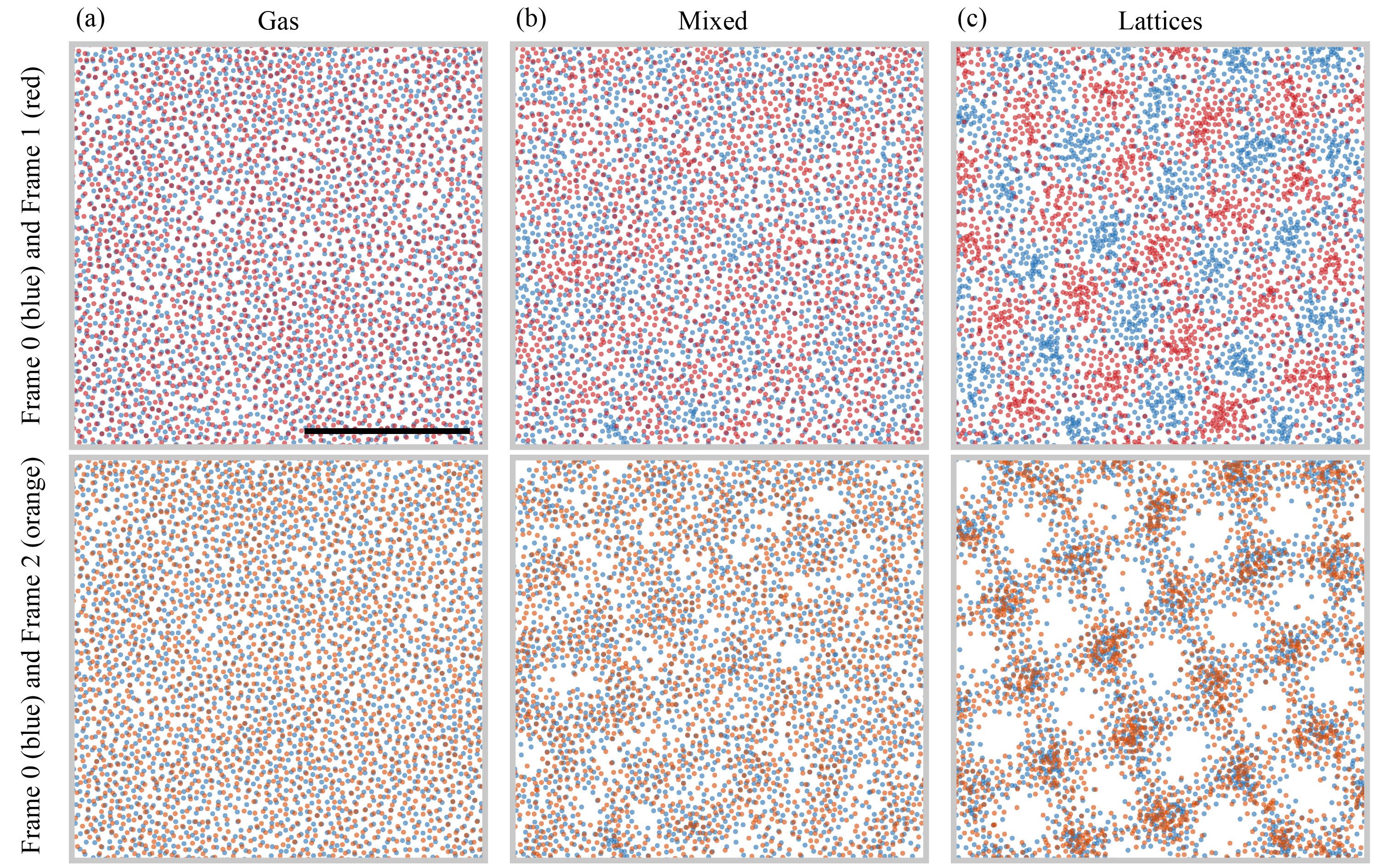}
\caption{
	Overlaps of two adjacent stable frames (top) and two every other stable frames (bottom) near the phase boundary between Gas and Lattices phases. $\tau_\text{on}$ are 30 ms (a), 40 ms (b) and 50 ms (c), respectively. The particle area fraction $\phi$ is 0.115. Particles are marked as colored dots. The scale bar is 0.2 mm.
}
\label{FigS12}
\end{figure}

\begin{figure}
\centering
\includegraphics[width=1.0\textwidth]{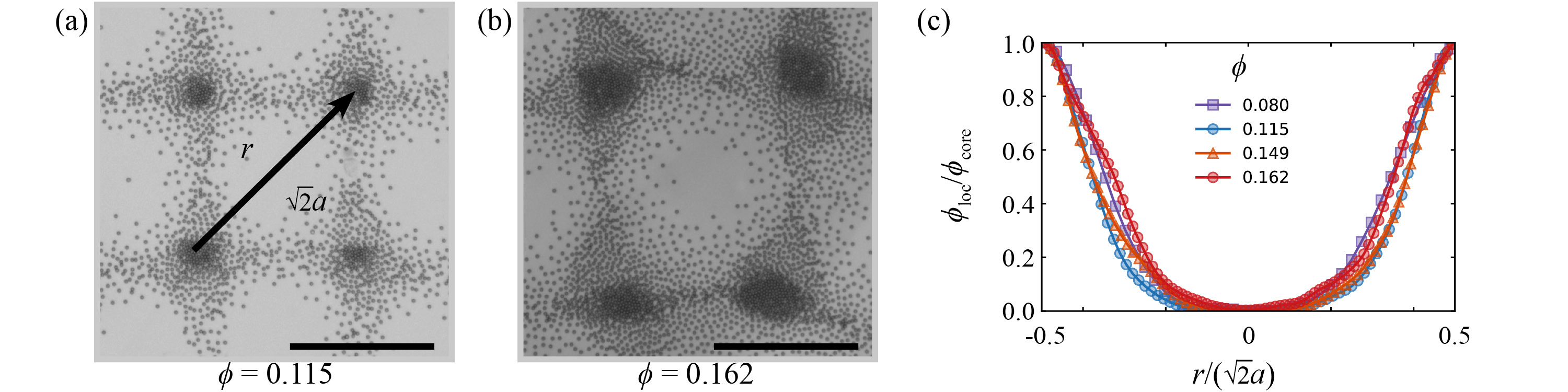}
\caption{
	Typical density profiles across the diagonal of square lattices at different average area fractions. $\phi_\text{loc}$ and $\phi_\text{core}$ are the local area fraction and the highest area fraction at lattice cores, respectively. $\tau_\text{on}$ is 100 ms. The scale bars are 0.2 mm in both (a) and (b). All curves show similar trends, indicating the similar density profiles at different average area fractions. 
}
\label{FigS13}
\end{figure}

\clearpage

\section*{Note S1}
\subsection*{Simulation details of the phenomenological model}
We develop a simple phenomenological model to confirm the formation of square lattices as a result of two main interaction mechanisms: alignment and repulsion. The model is similar to the one previously suggested in Ref.~\cite{bricard2015emergent} with a few simplifications and modifications. The behavior of the system is described in terms of particles coordinates $\mathbf{r}(k)$, velocities $\mathbf{V}(k)$ and hydrodynamic flow $\mathbf{V}_\text{h}(\mathbf{r})$, where $k=1..N$ is a particle index. Here, we need to emphasize that $\mathbf{V}_\text{h}$ is not necessarily a realistic hydrodynamic flow obtained by solving three dimensional Navier-Stokes equation, but a ``tool'' flow introduced to describe far-field particle interactions, mainly alignment. We have tested several types of repulsive potentials, including exponential and polynomial. The qualitative results of the simulation does not change significantly. 
In our model, the dimensionless particle velocity $\mathbf{V}_\text{p}$ is calculated as 
\begin{equation}
\mathbf{V}_\text{p}(k)= \mathbf{V}_0(k) -\nabla \sum_{k' \neq k} U_{k'}(\mathbf{r}(k))-\nabla U_\text{w} (\mathbf{r}(k)) +\alpha_\text{hp} \mathbf{V}_\text{h}(\mathbf{r}(k)) +  \mathbf{\xi},
\label{e1}
\end{equation}
where  $\mathbf{V}_0(k)$ is a unit vector parallel to the orientation of the $k$-th particle, $U_{k'}$ is a repulsive potential created by $k'$-th particle, $U_\text{w} \sim \tanh (\frac{r-R}{d})$ is a repulsive potential of a wall, $\alpha_\text{hp}$ reflects the sensitivity of the particle to the hydrodynamic flow $\mathbf{V}_\text{h}$ created by other particles, and $\xi$ is an uncorrelated white noise.

For better matching of collective dynamics with experimental observations, the hydrodynamic flow $\mathbf{V}_\text{h}$ has a non-negligible (while small) inertia.
\begin{equation}
\dot{\mathbf{V}}_\text{h}(\mathbf{r})=\alpha_1 \sum_k \mathbf{V}_\text{p} e^{-\frac{(\mathbf{r}(k)-\mathbf{r})^2}{\sigma_1^2} } + \alpha_2 \sum \nabla \varphi(k)  - \frac{\mathbf{V}_\text{h}(\mathbf{r})}{\tau_\text{h}}.
\label{e2}
\end{equation}
The first term in Eq \ref{e2} introduces the alignment of particles through hydrodynamic interactions accounted by the third term in Eq \ref{e1}. The meaning of this term can be understood as following. Each particle creates a gaussian-shape flow around it and this ``virtual'' flow aligns other particles in its vicinity. Since each particle creates a complex 3D dimensional flow, we ignored the violation of incomprehensibility condition $\nabla \cdot \mathbf{V}_h = 0$ in our 2D model introduced by this term. The second term in Eq \ref{e2} describes a potential flow around a moving sphere, where $\varphi(k) \sim 1/r^2 \text{cos}(\theta)$ is velocity potential created by $k$-th particle. According to our simulations this term is not required to visually reproduce the dynamics observed in the experiments and the lattices formation. The third term in Eq \ref{e2} describes dissipation of the flows due to the friction with the walls. 

To model the temporal variation of the system activity, particle velocities and hydrodynamic flows are reset to zero at the end of each cycle. The coordinates of particles remained unmodified until the reactivation of the system. Then particles start to move, interact with each other, and self-organize into lattices. Importantly, the initial direction of motion for each particle is parallel to the net repulsive force immediately upon system activation. If these directions are randomized, no stable lattices are observed.

\clearpage

\section*{Note S2}
\subsection*{Minimal number of rollers per unit cell}

According to the definition of the area fraction $\phi$,
\begin{equation}
	\phi = n \pi d^2/(4a^2), 
	\label{phi}
\end{equation}
where $n$ is the number of particles in a unit cell; $d$ is the particle diameter; $a$ is the lattice constant of the unit cell. 

From the dash line of the superposition of curves (Fig. 2c \Bo{inset})
\begin{equation}
	(a/d)\phi^{-0.5} = k\tau_\text{on}, 
	\label{k}
\end{equation}
where $\tau_\text{on}$ is the field-on time in ms; $k$ is the slope of the dash line. $k=$ 1.63 ms$^{-1}$. 

Combining Eq.~\ref{phi} and \ref{k}, 
\begin{equation}
	\phi = (\pi n)^{0.5}(2k)^{-1} \cdot \tau_\text{on}^{-1}. 
	\label{phi2}
\end{equation}

From the dash line of the phase boundary in Fig. 2b, the lattices exists when 
\begin{equation}
	\phi > A \cdot \tau_\text{on}^{-1}. 
	\label{phase}
\end{equation}
where $A= $ 5.5 ms. 
 
Combining Eq.~\ref{phi2} and \ref{phase}, finally we can have the critical number of rollers in a unit cell,  
\begin{equation}
	(\pi n)^{0.5}(2k)^{-1} > A. 
	\label{n}
\end{equation}
Or, 
\begin{equation}
	n > (2kA)^2/\pi \approx 100. 
	\label{n2}
\end{equation}
As a result, regardless of the area fraction of rollers, a minimum amount of 100 rollers per unit cell is needed to assemble the core-shell structure of reconfigurable stable lattices as shown in Fig. \ref{FigS6}. 

\clearpage

\bibliographystyle{naturemag}
\bibliography{Refs_abbrev}

\clearpage

\section*{Videos}

\subsection*{Video S1}
Square lattices in experiments. 
The electric field strength $E =$ 3.0 V $\SI{}{\micro\meter}$$^{-1}$. The frequency is 8 Hz and the duty cycle is 90 \%. $\tau_\text{on}$ and $\tau_\text{off}$ are 112.5 ms and 12.5 ms, respectively. The particle area fraction $\phi$ = 0.114. The frame size is 1.5 mm by 1.5 mm. The system is a cylindrical well with $D=$ 1 cm. The movie is 0.12X as the real speed.

\subsection*{Video S2}
Square lattices in simulations.
$\tau_\text{on}$ and $\tau_\text{off}$ are 37 and 3, respectively. There are 4000 particles in a square system with a size of 600 by 600.

\subsection*{Video S3}
Different phases.
\subsubsection*{Part 1}
Vortex phase. 
The electric field strength $E =$ 2.7 V $\SI{}{\micro\meter}$$^{-1}$. The frequency is 50 Hz and the duty cycle is 99 \%. $\tau_\text{on}$ and $\tau_\text{off}$ are 19.8 ms and 0.2 ms, respectively. The particle area fraction $\phi$ = 0.120. The frame size is 4.4 mm by 3.3 mm. The system is a cylindrical well with $D=$ 1 cm. The movie is 0.3X as the real speed. 

\subsubsection*{Part 2}
Domains phase. 
The electric field strength $E =$ 2.7 V $\SI{}{\micro\meter}$$^{-1}$. The frequency is 0.5 Hz and the duty cycle is 90 \%. $\tau_\text{on}$ and $\tau_\text{off}$ are 1800 ms and 200 ms, respectively. The particle area fraction $\phi$ = 0.120. The frame size is 4.4 mm by 3.3 mm. The system is a cylindrical well with $D=$ 1 cm. The movie is 1X as the real speed. 

\subsubsection*{Part 3}
Lattices phase. 
The electric field strength $E =$ 2.7 V $\SI{}{\micro\meter}$$^{-1}$. The frequency is 10 Hz and the duty cycle is 60 \%. $\tau_\text{on}$ and $\tau_\text{off}$ are 60 ms and 40 ms, respectively. The particle area fraction $\phi$ = 0.120. The frame size is 4.4 mm by 3.3 mm. The system is a cylindrical well with $D=$ 1 cm. The movie is 0.3X as the real speed. 

\subsubsection*{Part 4}
Flocks phase. 
The electric field strength $E =$ 2.7 V $\SI{}{\micro\meter}$$^{-1}$. The frequency is 50 Hz and the duty cycle is 70 \%. $\tau_\text{on}$ and $\tau_\text{off}$ are 14 ms and 6 ms, respectively. The particle area fraction $\phi$ = 0.120. The frame size is 4.4 mm by 3.3 mm. The system is a cylindrical well with $D=$ 1 cm. The movie is 0.3X as the real speed. 

\subsubsection*{Part 5}
Gas phase. 
The electric field strength $E =$ 2.7 V $\SI{}{\micro\meter}$$^{-1}$. The frequency is 10 Hz and the duty cycle is 20 \%. $\tau_\text{on}$ and $\tau_\text{off}$ are 20 ms and 80 ms, respectively. The particle area fraction $\phi$ = 0.120. The frame size is 0.6 mm by 0.6 mm. The system is a cylindrical well with $D=$ 1 cm. The movie is 0.3X as the real speed. 

\subsubsection*{Part 6}
Clusters phase. 
The electric field strength $E =$ 2.7 V $\SI{}{\micro\meter}$$^{-1}$. The frequency is 50 Hz and the duty cycle is 20 \%. $\tau_\text{on}$ and $\tau_\text{off}$ are 4 ms and 16 ms, respectively. The particle area fraction $\phi$ = 0.120. The frame size is 2.2 mm by 1.6 mm. The system is a cylindrical well with $D=$ 1 cm. The movie is 1X as the real speed.

\subsection*{Video S4}
Time evolution of square lattices in experiments.
The electric field strength $E =$ 3.0 V $\SI{}{\micro\meter}$$^{-1}$. The frequency is 8 Hz and the duty cycle is 90 \%. $\tau_\text{on}$ and $\tau_\text{off}$ are 112.5 ms and 12.5 ms, respectively. The particle area fraction $\phi$ = 0.114. The frame size is 4.4 mm by 3.3 mm. The system is a cylindrical well with $D=$ 1 cm. The movie is 20X as the real speed.

\subsection*{Video S5}
Time evolution of square lattices in simulations.
$\tau_\text{on}$ and $\tau_\text{off}$ are 47 and 3, respectively. There are 16000 particles in a square system with a size of 1200 by 1200.

\subsection*{Video S6}
Hysteresis of square lattices in experiments.
The electric field strength $E =$ 2.7 V $\SI{}{\micro\meter}$$^{-1}$. The frequency is 5 Hz. The duty cycle increases from 40 \% to 80 \% then decreases to 40 \%. $\tau_\text{on}$ increases from 80 ms to 160 ms then decreases to 80 ms. The duty cycle increases or decrease 2 \% (4 ms) by every 5 s. The particle area fraction $\phi$ = 0.120. The frame size is 3.3 mm by 3.3 mm. The system is a cylindrical well with $D=$ 1 cm. The movie is 16X as the real speed.

\subsection*{Video S7}
Hysteresis of square lattices in simulations.
$\tau_\text{on}$ increases from 27 to 77 then decreases to 27. $\tau_\text{on}$ increases or decrease 1 by every 10 frames. There are 16000 particles in a square system with a size of 1200 by 1200.